\documentclass[aps,prl,preprint,groupedaddress,showpacs]{revtex4}
\usepackage{graphicx}
\usepackage{graphics}
\usepackage{psfig}	
\usepackage{bm}

\begin{document}

\title{Dark Matter--baryon segregation in the non--linear evolution of coupled
 Dark Energy models.}
\vglue .4truecm

\author{ Roberto Mainini}
\vglue .1truecm

\affiliation{Universit\`a di Milano Bicocca, Piazza della Scienza 3, 20126 Milano,
 Italy ~\&~ I.N.F.N., Sezione di Milano}
\date{\today}

\begin{abstract}
The growth and virialization of spherical {\it top--hat} fluctuations,
in coupled Dark Energy models, causes segregation between Dark Matter
(DM) and baryons, as the gravitational infall into the potential well
proceeds more slowly for the baryons than for DM. As a consequence,
after attaining their turn--around and before full virialization,
halos have outer layers rich of baryons. Accordingly, a natural
ambiguity exists on the definition of the virial density contrast.  In
fact, when the outer baryon layers infall onto the DM--richer core,
they carry with them DM materials outside the original fluctuation;
hence, no time exists when all materials originally belonging to the
fluctuation --$\, $and only them$\, $-- have virialized.  Baryon--DM
segregation can have various astrophysical consequences on different
length--scales. The smallest halos may loose up to 50$\, \%$ of the
original baryonic contents and become hardly visible. Subhalos in
cluster--size halos may loose much baryonic materials, which could
then be observed as intra--cluster light. Isolated halos, in general,
can be expected to have a baryon component richer than the
cosmological proportions, due to the cosmic enrichement of baryons
lost in small halo encounters.
\end{abstract}

\pacs{98.80.-k, 98.65.-r }
\maketitle
\section{Introduction}
\label{sec:intro}
Dark Energy (DE) was first required by SNIa data, but a {\it flat}
cosmology with $\Omega_{m} \simeq 0.27$, $h \simeq 0.7$ and
$\Omega_{b} \simeq 0.045$ was soon favored by CMB and LSS observations
($\Omega _{m,b}$: matter, baryon density parameters; $h$: Hubble
parameter in units of 100 km/s/Mpc; CMB: cosmic microwave background
radiation; LSS: large scale structure). The nature of Dark Energy
is one of the main puzzles of cosmology. If it arises from false
vacuum, its pressure/energy--density ratio $w = p_{DE}/\rho_{DE}$ is
-1. For vacuum to yield DE, however, a severe fine tuning at the end
of the electroweak transition is required. Otherwise, DE could be a
scalar field $\phi$ self--interacting through a potential $V(\phi)$
\cite{sugra}, \cite{RP}, so yielding
\begin{equation}
\rho_{DE}=\rho_{k,DE}+\rho_{p,DE} \equiv {{\dot{\phi }}^{2}/2a^{2}}+V(\phi ),
~~~~~~~~~
p_{DE}=\rho_{k,DE}-\rho_{p,DE}
\end{equation}
provided that $\rho _{k}/V\ll 1/2$; then $-1/3 \gg w>-1$. Here, the
background metric reads $ds^{2} = a^{2} (-d\tau^{2}+ dx_i dx^i ) $
($i=1,..,3$) and dots yield differentiation with respect to the
conformal time $\tau $. This kind of DE is dubbed {\it dynamical} DE
(dDE) or {\it quintessence} and a large deal of work has been done on
it (see, e.g., \cite{PR} and references therein).

At variance from Dark Matter (DM), DE fluctuations tend to fade on
scales significantly smaller than the horizon. Furthermore, while
non--gravitational interactions between baryons and DE can be safely
excluded by experimental data, constraints on DM--DE interaction --~as
on most DM properties~-- can only be derived from cosmological
observations. In turn, such interaction could ease the cosmic
coincidence problem \cite{Amendola1999}, i.e. that DM and DE
densities, after being different by orders of magnitude for most of
the cosmic history, approach equal values in today's Universe. In a
number of papers, constraints on coupling, coming from CBR and LSS
data, were discussed \cite{Amendola1999,lucaclaudia,maccio}.

In this note we focus on coupled DE (cDE) models and study the growth
of a spherical {\it top--hat} fluctuation. This is done assuming a
SUGRA  \cite{brax} or a RP \cite{RP} self--interaction potential:
\begin{equation}
SUGRA ~~~~~~~~~~~~~~~~~~~~~~~~~~~~~~
V(\phi) = (\Lambda^{\alpha+4}/\phi^\alpha) \exp(4 \pi \phi^2/m_p^2)
\end{equation}
\begin{equation}
RP ~~~~~~~~~~~~~~~~~~~~~~~~~~~~~~~~~~~~~~~~~~~~~~~~~~
V(\phi) = (\Lambda^{\alpha+4}/\phi^\alpha)
\end{equation}
($m_p=G^{-1/2}$: Planck mass) as self--interaction potential.
Once $\Omega_{DE}$ is assigned, either $\alpha$ or the energy scale
$\Lambda$ can be freely chosen.

The background continuity equations for DE and DM then read
\begin{equation}
\ddot{\phi} + 2(\dot a/a) \dot{\phi} + a^2 V_{,\phi} = 
\sqrt{16\pi G/3} \beta a^{2} \rho _{c}~, 
~~~~ \dot{\rho _{c}} + 3 (\dot a/a)\rho_{c}  =  
-\sqrt{16\pi G/3} \beta \rho _{c} \dot \phi~,
\label{backg}
\end{equation}
while radiation and baryons obey standard equations.  The interaction
between DE and DM is parametrized by $\beta$ which, all through this
paper, is assumed constant and, in particular, not to depend on
$\phi$. Using WMAP data on $T$ anisotropies, a limit $\beta < \sim
0.3$ has been established. A more stringent limit on $\beta$, found
for the potential \cite{RP} by studying halo profiles in cDE
simulations \cite{maccio}, does not apply to this case.

In this paper we explored different values of $\beta$, taking $\Lambda
= 10^3\, $GeV and $\Omega_m = 0.30$, $\Omega_b = 0.05$ (model I);
where significant, this was compared with a cosmology with $\Omega_m =
0.25$, $\Omega_b = 0.04$ (model II); $h=0.7$ was taken for both
models. Full detail will be given for model I with SUGRA potential.

This model provides a fair fit to CMB data, even performing slightly
better than $\Lambda$CDM \cite{cmb}. Model II and RP potential are
then considered just to explore how results depend on the choice of
cosmological parameters and potential.

Accordingly, results for Model I and SUGRA will be compared with
Model II and SUGRA and/or Model I and RP. Some quantitative outputs
will be however provided also for Model II and RP.

The evolution of a spherical {\it top--hat} fluctuation, in the linear
and non--linear regimes, can be found analytically in a standard CDM
model. Here the density contrast at virialization $\Delta_v \simeq
180$. Although assuming spherical symmetry, which is not the geometry
of growing fluctuations, this treatment, associated with a Press \&
Schechter \cite{PS} --~or similar \cite{ST}~-- formulation, predicts
observational and numerical mass functions with a high degree of
approximation. The value of $\Delta_v$ is also used to extract
virialized systems from galaxy samples and n--body simulations.

The analysis of the spherically simmetric growth has been extended to
$\Lambda$CDM models \cite{Lahav_brian} and other
models with uncoupled DE, where it was used with similar success \cite{io,collapse1}.
In these cosmologies, the simple
system of equations yielding spherical evolution requires a numerical
solution.

Models with DM-DE coupling are also considered by \cite{collapse2} but they do not
take into account the different dynamics between coupled and uncoupled components 
as well as their bias (see eq.(\ref{bias}) below).

Infact, in the case of cDE, the dynamics of baryons and DM are
different.  An {\it effective} Newtonian theory can be safely used to
describe their physical evolution, when radiative materials are
dynamically irrelevant and for scales well below the horizon scale,
both for small and non--linear fluctuations, as was done, e.g., by
\cite{maccio}. 

In the newtonian limit, the DM--DE interaction causes: (i) DM particle
masses to vary $\propto \exp(-C \phi)$; (ii) the gravitational
constant between DM particles to become $G^* = \gamma G$. Here
\begin{equation}
C = \sqrt{16\pi G/3} ~\beta ~~,~~~~~~
\gamma = 1+4\beta^2/3~.
\label{Cgamma}
\end{equation}
Taking this into account, at the initial time $\tau_{i}$, we assume
that both DM and baryons have a {\it top--hat} fluctuation of
identical radius $R_{TH,i}$, expanding together with the Hubble flow.
The linear theory \cite{Amendola03} prescribes that fluctuation amplitudes in baryons
and DM start from a ratio
\begin{equation}
{\delta_{b} \over \delta_c }
\simeq
{ 3 \Omega_c \over
3 \gamma\, \Omega_c + 4\beta\, X \mu}~.
\label{bias}
\end{equation}
Here $\Omega_c$ is the DM density parameter, $\mu = (\dot
\delta_{c,b}/\delta_{c,b})/(\dot a/a)$, $X = \sqrt{4 \pi/3}\, \dot
\phi/(m_p \dot a/a)$.  Due to the different interaction strength, as
soon as a non--linear regime is reached, the size $R^{b}$ of the
baryon perturbation begins to exceed the size $R^{c}$ of the DM
perturbation. Hence, a part of the baryons initially within $R_{TH,i}$
leak out from $R^c$, so that DM does not feel the gravity of all
baryons, while baryons above $R^c$ feel the gravity also of
initially unperturbed DM layers. As a consequence, above $R^c$,
the baryon component deviates from a {\it top--hat} geometry, while a
secondary perturbation in DM arises also above $R^c$ itself.

After reaching maximum expansion, contraction will also start at
different times for different components and layers. Then, inner
layers will approach virialization before outer layers, whose later
fall--out shall however perturb their virial equilibrium.  This
already outlines that the onset of virial equilibrium is a complex
process.

Furthermore, when the external baryon layers fall--out onto the
virialized core, richer of DM, they are accompanied by DM materials
originally outside the top--hat, perturbed by baryon over--expansion.

The time when the greatest amount of materials, originally belonging
to the fluctuation, are in virial equilibrium occurs when the DM
top--hat has virialized, together with the baryon fraction still below
$R^c$. A large deal of baryonic materials are then still falling
out. But, when they will accrete onto the DM--richer core, they will
not be alone, carrying with them originally alien materials.  There
will be no discontinuity in the fall--out process when all original
baryons are back. The infall of outer materials just attains then a
steady rate.


In order to follow the dynamical evolution of a systems where each
layer feels a substance--dependent force, a set of concentric shells,
granting a sufficient radial resolution, needs to be considered. In
the next section we shall provide the equations of motion for
Lagrangian shell radii and discuss the whole expansion and
recontraction dynamics. More details on the virialization process are
provided in Section 3. Section 4 has a technical content and is
devoted to working out the Newtonian regime from General Relativistic
equations. In Section 5 a brief discussion on the relevance of
baryon--DM segregation, on various length--scales, is given. Final
conclusions are drawn in Section 6.

\section{Time evolution of concentric shells}

In order to write dynamical equations, we shall make use of the
comoving radii $ b_n = R^b_n/a $ and $ c_n = R^c_n/a $
for the $n$--th baryon or DM shell, respectively.  The masses of
layers depend on the $R_n$ distribution at the initial time. Let
$M_n^{b,c}$ be the mass of the $n$--th layer for each component and
let $\bar M_{n}^{b,c}$ be the mass for the same layer and component in
the absence of the perturbation.  Accordingly, we can define the mass
excess $\delta M_n^{b,c} = M_{n}^{b,c} - \bar M_{n}^{b,c}$, for the
$n$--th layer.
While the masses of baryonic shells keep constant while
they expand and then recontract, according to eq.~(\ref{backg})
it is
\begin{equation}
 M_n^c (\tau) = M_n^c (\tau_{in}) \exp\{-C
[\phi(\tau)-\phi(\tau_{in})]\}~.
\label{mvar}
\end{equation}

Similarly, let  $\Delta M^{b,c}(<r)=  M^{b,c}(<r)- \bar M _{b.c}(<r)$ be the mass excess 
within the physical radius $R=ar$ and let then be
\begin{equation}
\Delta M_n^b = \Delta M^{b}(<b_n) + \Delta M^{c}(<b_n)~,~~~
\Delta M_n^c = \Delta M^{b}(<c_n) + \gamma \Delta M^{c}(<c_n)~,
\label{dm}
\end{equation}
taking into account both the mass variation (eq.~\ref{mvar}) and that
$G^* \neq G$. Here $\Delta M^{b,c}(<r)$ are masses, approximately
given by the sum of all the $\delta M_n^{b,c} ~$ for which
$b_n,c_n<r$. $\Delta M^{b,c}_n$, instead, are gravity sources, given by
masses corrected --$\, $somewhere$\, $-- by suitable factors,
different for DM and baryons.

Problems due to the systematic discrepancy between the
comoving radii $b_n$ and $c_n$ can be cured by increasing the
resolution and taking into account only the fraction of the {\it last}
shell below the radius considered. The shell dynamics can then
be described through the equations
\begin{equation}
\ddot c_n = - \left( {\dot a \over  a} - C\dot \phi \right) \dot c_n
- G {\Delta M_n^c \over a c_n^2}~,~~~
\ddot b_n = - {\dot a \over a} \dot b_n
- G {\Delta M_n^b \over a b_n^2}~,
\label{shells}
\end{equation}
which 
are derived in Sec.~4.
These equations are to be integrated together with the first of
eqs.~(\ref{backg}) and the Friedman equation.

The eqs.~(\ref{shells}) are analogous for DM and baryons.  Until $b_n
\sim c_n$, it is $\Delta M_n^{b} < \Delta M_n^{c}$ (eq.~\ref{dm}), for
any $\beta > 0$. Hence, the gravitational push felt by DM layers is
stronger. The extra term $C\dot \phi \dot c_n$ adds to this push.  In
fact, the comoving variable $ c_n$ has however a negative derivative.
As a consequence, the DM comoving radius $c_n$ decreases more rapidly
than the baryon comoving radius $b_n$ (or $R_n^{c}$ increases more
slowly than $R_n^{b}$) and the $n$--th baryon radius may gradually
exceed the $(n+1)$--th DM shell, etc.$\, $. As a consequence, the sign
of $\Delta M_n^{b} - \Delta M_n^{c}$ can even invert and one can
evaluate for which value of $\phi-\phi_i$ this occurs. Once
$\phi(\tau)$ is known, also the time when this occurs can be found.
Until then, however, DM fluctuations expand more slowly and mostly
reach their turn--around point earlier, while baryons gradually leak
out from the fluctuation bulk.

\begin{figure}
\begin{center}
\includegraphics*[width=12cm]{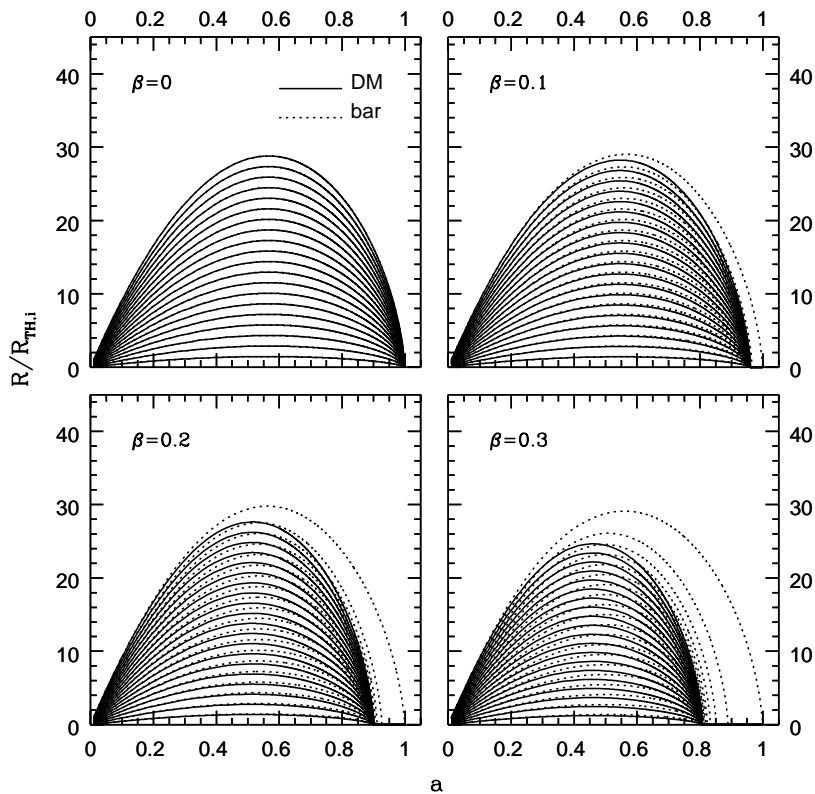}
\end{center}
\vskip -1.2truecm
\caption{Evolution of a sample of baryon and DM layer radii, extending
up to the top--hat radius, in model I, for couplings $\beta$ up to
0.3~.}
\label{recol}
\end{figure}
\begin{figure}
\begin{center}
\includegraphics*[width=10cm]{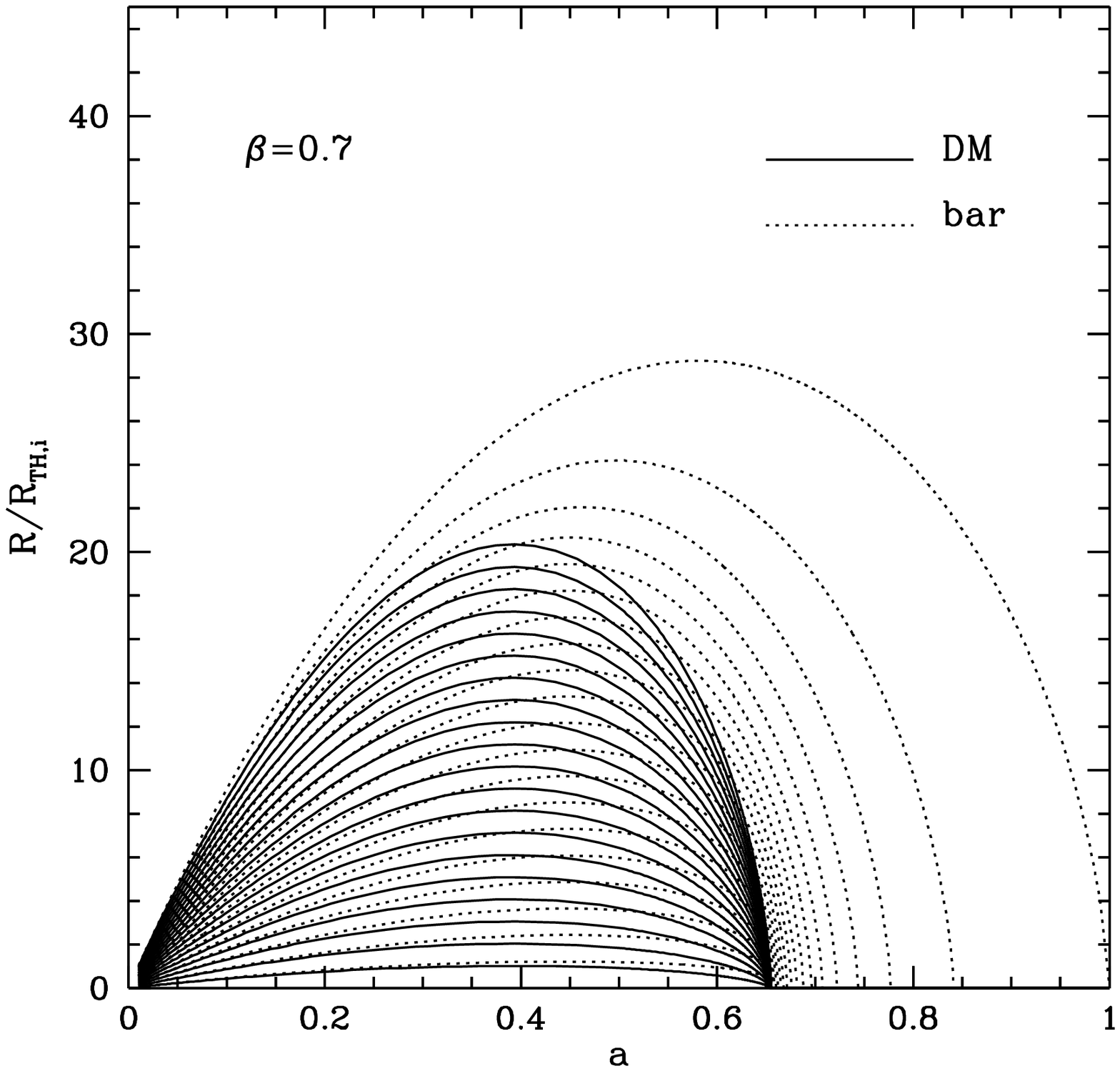}
\end{center}
\caption{Evolution of a sample of baryon and DM layer radii in model
I, for $\beta = 0.7$~. Although such coupling is somehow large,
the plot allows the reader to see the detail of layer evolution.}
\label{recol7}
\end{figure}

This is visible in the Figure \ref{recol}, for model I.  Here the time
dependence of a sample of radii $R_n^{b,c}$ is shown. The greatest
radii shown are the top--hat radii. Solid and dotted lines yield the
$a$ dependence of the ratios $R_n^c/ R_i$ and $R_n^b/
R_i$ respectively, for $\beta = 0,\, 0.1,\, 0.2,\, 0.3$~. These
plots are obtained from the solution of eqs.~(\ref{shells}) down to
the collapse of the last baryon shell, assumed to occur at $a=1$
(today). An extreme case ($\beta = 0.7$) is also shown in
Fig.~\ref{recol7}. The actual number of radii $R_n^{b,c}$ used depends
on the level of precision wanted and radii do not need to be
equi--spaced.  The requirement fulfilled here is that, doubling the
radius number in the critical regions, where profiles are expected to
suffer a stronger deformation, causes a shift in all outputs never
exceeding 1$\, \%$. To this aim, up to $\sim 1000$ radii had to be
used. If more precision is wanted, it can be easily achieved at the
expenses of using a greater computer time. With a single processor 1.5
Gflops PC, the numerical program takes $\sim 90$ minutes to run a
single case, with $\sim 1000$ radii.

\begin{figure}
\begin{center}
\includegraphics*[width=9cm]{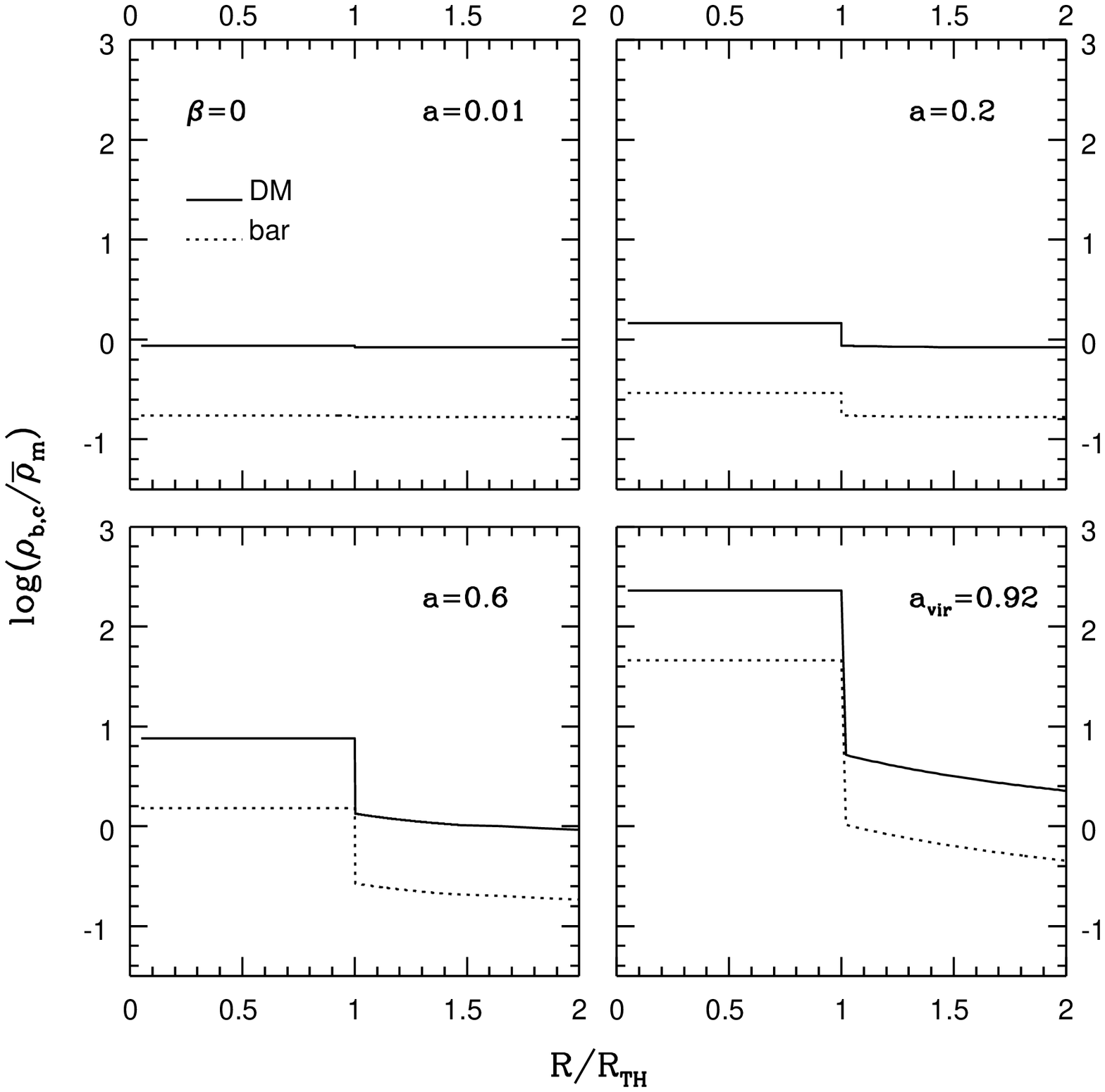}
\end{center}
\caption{Density profiles at different $a$ values for uncoupled DE
model. Solid (dotted) lines refer to DM (baryons), In the
absence of coupling the two profiles have the same shape.
The shell treatment, in this case, in unnecessary; notice,
however, the density increase {\it outside} the {\it top--hat}.~}
\label{beta0}
\end{figure}

\begin{figure}
\begin{center}
\includegraphics*[width=9cm]{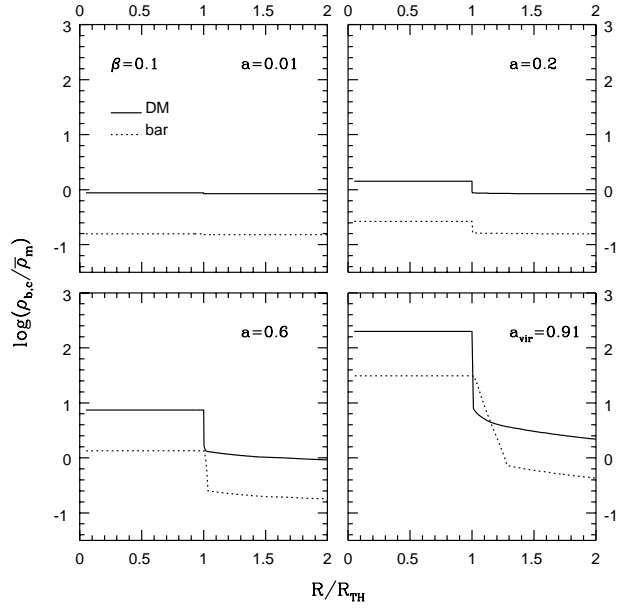}
\end{center}
\caption{Density profiles at different $a$ values for $\beta = 0.1$
model. Notice the progressive deformation of the baryon profile
(dotted lines) in respect of the DM profile (solid line).~}
\label{beta1}
\end{figure}

\begin{figure}
\begin{center}
\includegraphics*[width=9cm]{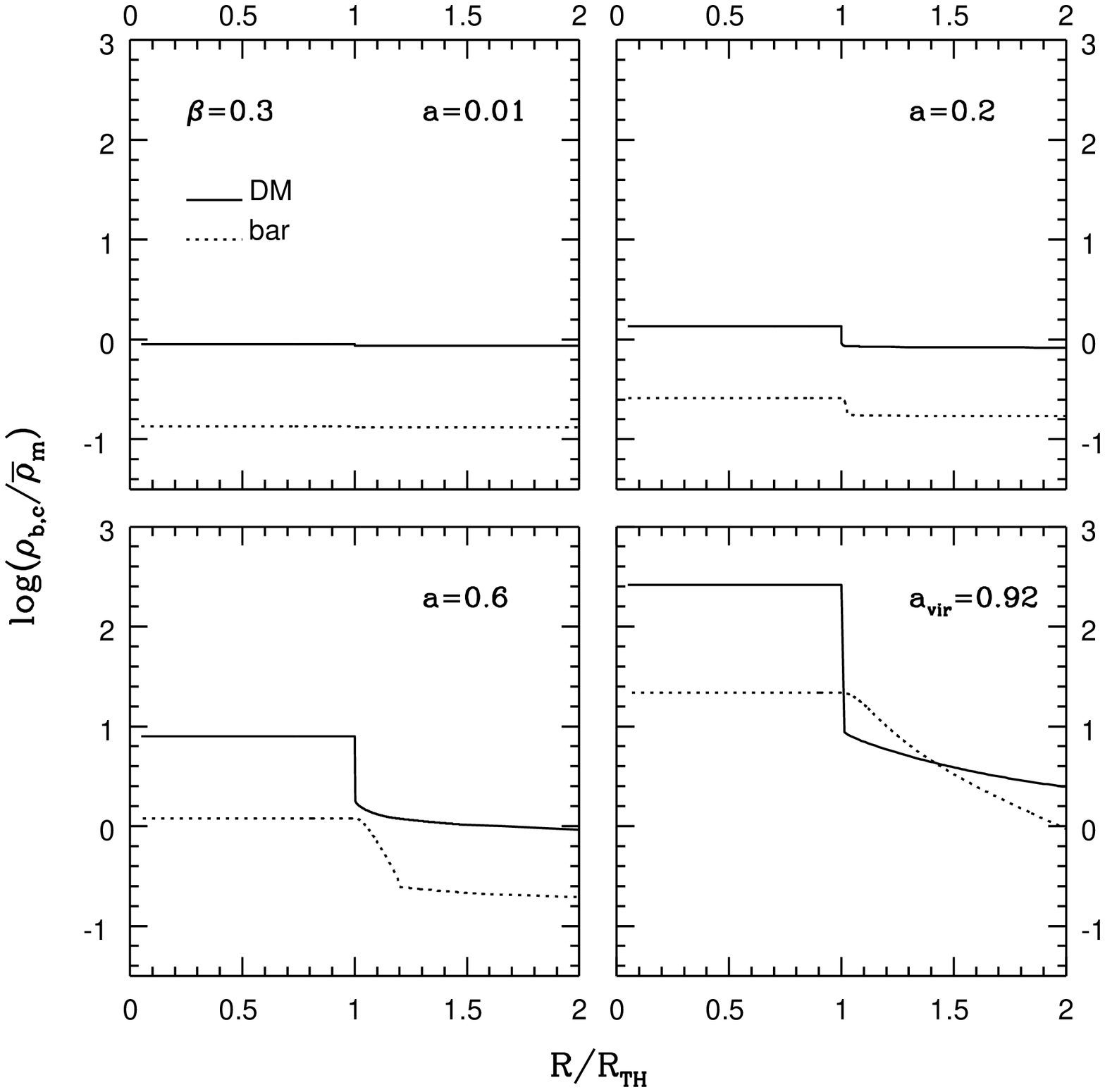}
\end{center}
\vskip -1.2truecm
\caption{Density profiles at different $a$ values for $\beta = 0.3$
model. The distorsion of the baryon profile (dotted line), in this
case, is already noticeable at turn--around and, around virialization,
the baryon profile has almost lost its initial top--hat shape.~}
\label{beta2}
\end{figure}

Full collapse is not expected in the physical case, as the $R_n$
decrease shall stop when virialization is attained and, in the next
section, the virialization condition will be discussed.  Figures
\ref{beta0}, \ref{beta1}, \ref{beta2}, however, assume that the
spherical growth stops when all DM originally in the top--hat, and the
baryons kept inside it, virialize.

In Figure \ref{beta0}, the effects of the spherical growth are shown
in the limit of no coupling ($\beta = 0$) for model I. This plots
allows to outline an effect that is usually disregarded, i.e., that
also materials outside from the top--hat are perturbed by its
growth. Their density, in fact, increases in respect to average. As
shown in Figure \ref{recol}, the turn--around occurs for $a \sim
0.6$. At such time, the density enhancement of exterior layers extends
up to a radius approximately double of the top--hat radius. When the
virialization radius is reached by DM and inner baryons ($a \simeq
0.92$), the perturbed sphere is even wider and, at twice the top--hat
radius, the density enhancement is $\sim 2.4$. These effects, however,
do not bear a great relevance in the $\beta=0$ case, where layer
sharing is unessential, as growth is self--similar over all shells
inside the top--hat.

In Figure \ref{beta1}, the case $\beta = 0.1$ is considered. Already
at $a = 0.2$, well before the turn--around, the slope of the top--hat
boundary, for baryons, is no longer vertical.  At $a=0.6$
(approximately turn--around), not only the baryon boundary is bent,
but a similar effect is visible also for DM, where the density
increase at $R>R_{TH}$, is however greater than for $\beta=0$.
The effect is even more pronounced at $a_{vir } \simeq 0.92$.
Figure \ref{beta2} refers to $\beta = 0.3$, where all above
effects are present and stronger. 

The increased density of DM shells outside the top--hat is relevant,
here, because it fastens the recollapse of outer baryons. In turn, the
enhanced baryon density, outside from DM top--hat, modifies the
dynamics of external DM layers, as well.

Let us also outline that the initial fluctuation amplitude is different
in Figures~\ref{recol} and \ref{beta0}, \ref{beta1}, \ref{beta2}.
In the former case, it is tuned so that the outest layer fully
recollapses at $a=1$. In the latter case, instead, only the DM
fluctuation would have fully recollapsed at $a=1$.

Another point which ought to be noticed in the same Figures is that
the only density discontinuity occurs at the boundary of the DM
top--hat and is due to the DM component only. Baryon density begin to
decline there, but is not dicontinuous. The baryon excess, due to the
initial top--hat, makes them intially denser than DM at $R>
R^c$. The density ratio inverts at some outer radius and the
baryon density decrease is faster than for outer DM layers. The radius
$R^b$, where the most external baryon top--hat shell lays, is
visible in most plots, as a discontinuity of the density slope. It
however lays even below $2\, R^c$, in some profiles.

\begin{figure}
\begin{center}
\vskip -.9truecm
\includegraphics*[width=9cm]{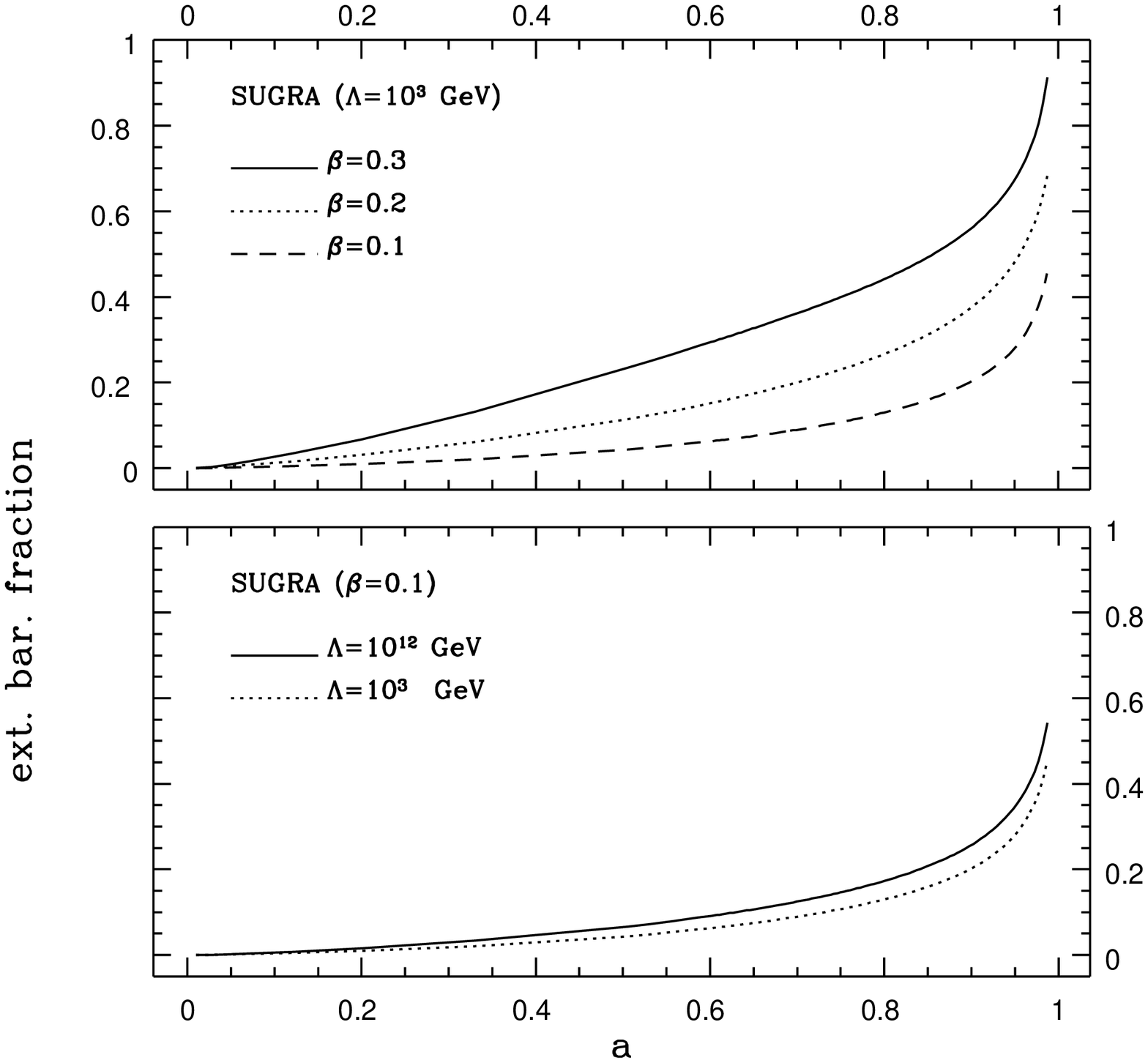}
\end{center}
\vskip -1.2truecm
\caption{Baryon fraction outside top--hat approaching virialization (SUGRA model).~}
\label{outbeta1}
\end{figure}

\begin{figure}
\begin{center}
\includegraphics*[width=9cm]{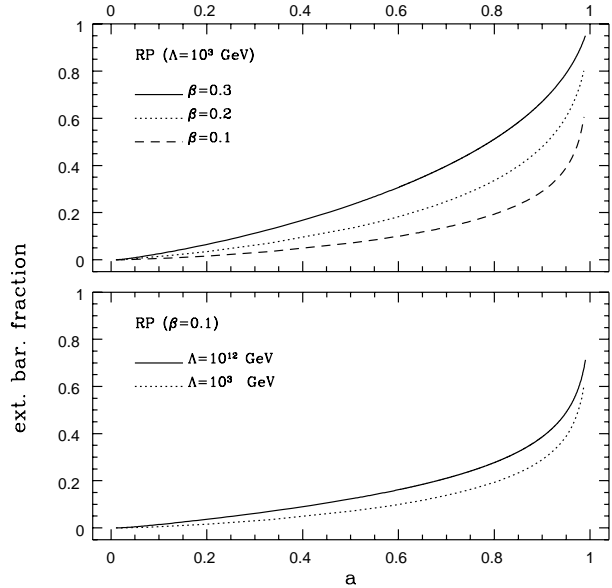}
\end{center}
\vskip -1.2truecm
\caption{Baryon fraction outside top--hat approaching virialization (RP model).~}
\label{outbeta2}
\end{figure}

As outlined before, the most apparent physical effect is the outflow
of baryon layers from the DM top--hat. The outflown baryon fraction
increases with $a$. In the top panel of  Figure \ref{outbeta1}, such fraction is shown
for $\beta$ varying from 0.1 to 0.3$\, $ in the case of SUGRA potential. 
For $a \sim 0.92$, i.e. when DM and inner baryons have attained their virialization
radius, the fraction of baryons which have leaked out from the
fluctuation is $\sim 20\, \%$ for $\beta = 0.1$ and reaches $\sim 58\,
\%$ for $\beta = 0.3\, $. These  values increase by an additional  $ 10\, \%$ in RP case.
This is shown in the top panel of Figure \ref{outbeta2}. 
The lower panel of both Figs. \ref{outbeta1} and  \ref{outbeta2} shows 
that the effect depends mildly on the scale $\Lambda$ and increases when $\Lambda$ is greater.
\section{Virialization}
As already outlined, in Figures \ref{beta1}, \ref{beta2} the DM
profile preserves its initial top--hat configuration which, on the
contrary, is gradually distorted in the baryon component. In order to
set the virialization time, therefore, one must first state to which
shells it refers. An obvious option amounts to relating virialization
to top--hat materials which are not polluted by accretion of external
layers. This definition (option I), however, leaves out a large deal
of baryons, which will accrete later on, although accompanied by alien
DM. Another choice (option II) is then to wait until all baryons
are allowed to virialize, regardless of the intruding DM, which
alterates the original proportions. Any intermediate choice is also
legitimate. For instance, one could refer to the time when the
total mass reaching its virialization radius equates the total
mass intially inside the top--hat, regardless of the DM/baryon ratio,
which will be however finally increased.

Most of this discussion does however risk to be non--physical. The
study of the evolution of a spherical top--hat is significant when
applied to a realistic cosmic environment. We shall take this point of
view in Section 5. Here below we shall define that {\it virialization}
has occurred when DM and inner baryons would have fully recollapsed
(option I).

During fluctuation evolution, the potential energies both of DM and
baryons will be made of three terms, arising by self--interaction,
mutual interaction, and interaction with the DE field:
\begin{equation}
U^c(R) = U^{cc}(R)+ U^{cb}(R)+ U^{c,DE}(R) =
4 \pi \int^R_0 dr ~r^2\, \rho_c(r)~ [\bar \Psi_c(r) + \Psi_b(r) + \Psi_{DE}(r)]
\label{uc}
\end{equation}
\begin{equation}
U^b(R) = U^{bb}(R)+ U^{bc}(R)+ U^{b,DE}(R) =
4 \pi \int^R_0 dr ~r^2\, \rho_b(r)~ [\Psi_b(r) + \Psi_c(r) + \Psi_{DE}(r)]
\label{ub}
\end{equation}
Here the sum on the layers is replaced by a volume integral, extended
to a sphere of radius $R$. While 
\begin{equation}
\Psi_b (r) = -{4 \pi \over 3} G \rho_{b}(r) r^2~,~~
\Psi_{DE} (r) = -{4 \pi \over 3} G \rho_{DE}(r) r^2~,~~
\end{equation}
in both expressions, a subtle difference exists for $\Psi_c$.
In $U^b$ we have simply
\begin{equation}
\Psi_c (r) =  -{4 \pi \over 3} G \rho_{c}(r) r^2~,~~
\end{equation}
as for the other components, but this expression is different in
$U^c$, where it reads
\begin{equation}
\bar \Psi_c (r) =  -{4 \pi \over 3} \{ \gamma G [\rho_{c} - \bar \rho_c (r)] 
+ G \bar \rho_{c} \}r^2~,
\label{diversa}
\end{equation}
$\bar \rho_c$ being the background DM density. The different dynamical
effect of background and DM fluctuation arises from the different ways
how its interaction with DE is treated. Energy exchanges between DM
and DE, for the background, are accounted for by the r.h.s. terms in
eqs.~(\ref{backg}). In this case no newtonian approximation was
possible and was made. For DM fluctuation, instead, the effects of
DM--DE exchanges are described by a correction to the gravitational
constant $G$, becoming $G^* = \gamma G$, which adds to the dependence
of DM density on $\phi$. This ought to be taken into account in the
fluctuation evolution, as is done in eq.~(\ref{diversa}).

Together with the potential energies, the kinetic energies
\begin{equation}
T^c(R)={1\over 2} \int_V dV \rho_c ~ \dot r^2~,~~ 
T^b(R)={1\over 2} \int_V dV \rho_b ~ \dot r^2
\label{kinetic}
\end{equation}
for DM and baryons within $R$, due to the collapse
motion, ought to be considered. The virialization condition then reads
\begin{equation}
2\, T(R) = R\, dU(R)/dR
\label{viri}
\end{equation}
Here $T(R) = T^c(R) + T^b(R)$ and $U(R) = U^c(R) + U^b(R)$. 
If $R$ is selected within the DM top--hat, $\rho_{b,c,DE}$ do not
depend on $r$  (such dependence would become important if virialization
were defined taking into account layers outside the DM top--hat) 
and then integrals (\ref{uc}), (\ref{ub}) and (\ref{kinetic})
can be performed obtaining
\begin{equation}
U^c(R)= -{3\over 5}~G~{M^c[\bar M^c + \gamma \Delta M^c + M^b] \over R}
- {4 \pi \over 5} M^c \rho_{DE} R^2
\end{equation}
\begin{equation}
U^b(R)= -{3\over 5}~G~{M^b [M^b +  M^c] \over R}
- {4 \pi \over 5} M^b \rho_{DE} R^2
\end{equation}
\begin{equation}
T^c(R)= {1\over 2}~{3\over 5}~M^c \dot R^2 
\end{equation}
where the relation $\dot r / r = \dot R / R$ is used to calculate $T^c(R)$.
Note that this relation is not valid for $T^b(R)$ because different
baryon layers have different growth rates. 
Kinetic energy for baryons is then obtained by 
\begin{equation}
T^b(R) = \sum_n T^b_n = \sum_n {1 \over 2} M^b_n (\dot R^b_n)^2 
\end{equation}
Here the sum is extended on all $R^b_n < R$.
Notice that, because of DE--DM interaction yielding DM particle mass
variation, we cannot require energy conservation for DM and baryons
between turn--around and virialization. However, energy conservation
in the spherical growth is violated, also in cosmologies with
uncoupled dynamical DE, once we assume full DE homogeneity. This problem
as well as the possibility to relax the homogeneity assumption for DE are discussed
by \cite{maor_wang}.

\begin{figure}
\begin{center}
\includegraphics*[width=11cm]{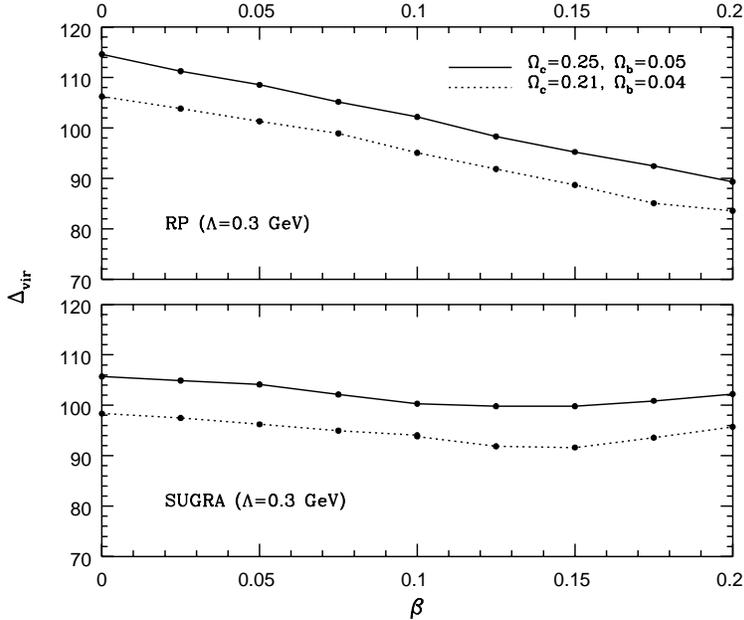}
\end{center}
\vskip -1.2truecm
\caption{Density contrast at virialization in cDE models.~}
\label{Dvir}
\end{figure}

In Figure \ref{Dvir} we report the $\beta$--dependence of the density
contrast at virialization, both for models I and II, both for SUGRA and RP potential.

\section{The Newtonian Regime}
This section debates the passage from the General Relativistic
treatment to the Newtonian Regime equations, for cDE theories. The
discussion partially resumes arguments made by \cite{Amendola03} and
\cite{maccio}, whose outcomes are used to work out
eqs.~(\ref{mvar}). The reader, who is not interested in deriving
these equations, can skip this section.

When fluctuations are considered, in the Newtonian gauge the metric
reads
\begin{equation}
ds^{2}=a^2(\tau)[-(1+2\psi )d\tau ^{2}+(1-2\psi )dx_{i}dx^{i}],
\end{equation}
$\psi $ being the gravitational potential. No anisotropic stresses are
considered here.
As usual, fluctuations are expanded in components of wavenumber $\bf k$
and let be $\lambda ={\mathcal{H}}/k$ (${\cal H} = \dot a/a$).
Dealing with fluctuations well after recombination, radiation is
neglected. Baryon (DM) density fluctuations are dubbed $\delta_{b}$
($\delta_{c}$); DE field fluctuations $\delta \phi $ and (DM or
baryon) velocity fields ${\bf v}_{c,b}$ are described by the variables
\begin{equation}
\varphi =\sqrt{\frac{4\pi G}{3}}\, \delta \phi ~,~~
\theta^{c,b}=i{\frac{\bf k \cdot {\bf v^{c,b}}}{{\cal H}}}.
\label{varphi}
\end{equation}
Without restriction on the DE potential $V(\phi$) one can define the functions
\begin{equation}
f = \phi^{-1} \sqrt{3/16\pi G} ~\ln(V/V_o)
,~~
f_{1}=\phi {\frac{df}{d\phi }}+f,~~ 
f_{2}= \phi {\frac{df}{d\phi }}+2f+f_{1}~;
\end{equation}
($V_o$ is a reference value of the potential);
it is also useful to define $X^{2} = 4\pi G \dot\phi^2 /3 {\cal
H}^{2}$ (as in the introduction) and $Y^{2}=8\pi GV(\phi)\, 
a^2 /3 {\mathcal{H}}^{2}$. 
The gravitational and $\varphi$ field equations then read
\begin{equation}
\psi = -{3 \over 2} \lambda^2 (\Omega_b \delta_b +
\Omega_c \delta_c + 6X \varphi + 2X \varphi'- 2 Y^2 f_1\, \varphi )~,~~
\psi' = 3 x \varphi - \psi ~,~~~~~~~~
\label{poten1}
\end{equation}
\begin{equation}
\varphi ''+\big (2+\frac{ { \cal H}^{'}}{ {\cal H}}\big )\varphi '+
\lambda ^{-2}\varphi -12X\varphi +4\psi X+2Y^{2}(f_{2}\varphi -f_{1}\psi )=
\beta \Omega _{c}(\delta _{c}+2\psi ),
\label{phi2}
\end{equation}
keeping just the lowest order terms in $\lambda$, to obtain the
Newtonian limit; if DE kinetic (and/or potential) energy substantially
contributes to the expansion source, $X$ (and/or $Y$) is
$\mathcal{O}$(1).

In the Newtonian limit, we must also neglect the derivatives of
$\varphi $, averaging out the oscillations of $\varphi $ and the
potential term $f_{2}Y^{2}\varphi $, requiring that $\lambda
<<(f_{2}Y)^{-1}$ (remind that $Y$ is $\mathcal{O} (1)$).  Furthermore,
in eq.~(\ref{phi2}), the metric potential $\psi $ ($\propto
\lambda^2$) can also be neglected. Accordingly, eq.~(\ref{poten1}) and
(\ref{phi2}) become
\begin{equation}
\psi =-\frac{3}{2}\lambda ^{2}(\Omega _{b}\delta _{b}+\Omega _{c}\delta _{c}),\
~,~~
\lambda ^{-2}\varphi \simeq \beta \Omega _{c}\delta _{c}.\label{approx2}
\label{approx1}
\end{equation}
(the former one is the Poisson equation). Then, from the
stress--energy {\it pseudo}--conservation $T_{\nu ;\mu }^{\mu }=0$,
one can derive a couple of equations telling us
how $\delta _{c,b}$ and $\theta _{c,b}$ depend on
$a$:
\begin{eqnarray}
\label{deltaII}
{\delta _{c}}\, '' & = & 
-{\delta _{c}}\, '\big (1+\frac{{\mathcal{H}}'}
{\mathcal{H}}-2\beta X\big )+\frac{3}{2}
(1+\frac{4}{3}\beta ^{2})\Omega _{c}\delta _{c}+
\frac{3}{2}\Omega _{b}\delta _{b},\label{delta1}\\
{\delta _{b}}\, '' & = & 
-{\delta _{b}}\, '\big (1+\frac{{\mathcal{H}}'}
{\mathcal{H}}\big )+\frac{3}{2}(\Omega _{c}\delta_{c}
+\Omega _{b}\delta _{b}),\label{delta2}
\end{eqnarray}
\begin{eqnarray}
{\theta _{c}}\, ' & = & 
-\theta _{c}\big (1+\frac{{\mathcal{H}}'}{\mathcal{H}}
-2\beta X\big )-\frac{3}{2}\big (1+\frac{4}{3}\beta^{2}\big)
\Omega _{c}\delta _{c}-
\frac{3}{2}\Omega _{b}\delta _{b},
\label{arr1},
\\
{\theta _{b}}\, ' & = & 
-\theta _{b}\big( 1+\frac{{\mathcal{H}}'}{\mathcal{H}} \big)
-\frac{3}{2}(\Omega _{c}\delta _{c}+\Omega _{b}\delta _{b})~;
\label{arr2}
\end{eqnarray}
here $'$ yields differentiation with respect to $\alpha = \ln a$.
Assuming $\Omega_b << \Omega_c$ and putting 
$\delta_c \propto e^{\int \mu(\alpha) d \alpha}$  and $\delta_b = b \delta_c$ 
with $b={\it cost}$ from 
the eqs.(\ref{delta1}),~(\ref{delta2}), we obtain the bias 
factor (\ref{bias}). 
The acceleration of a single DM or baryon particle of mass $m_{c,b}$ can be 
instead derived from eqs.~(\ref{arr1}),~(\ref{arr2}).  Let us set it in
the void, at a distance $r$ from the origin, where a DM (or baryon)
particle of mass $M_{c}$ (or $M_{b}$) is set, and let us remind that,
while the usual scaling $\bar \rho_b \propto a^{-3}$ holds, it is
\begin{equation}
\bar \rho_c = \bar \rho_{oc} a^{-3}
e^{-C (\phi -\phi_0)},~~ \rho_{M_c} = M_{oc} a^{-3} e^{-C (\phi -\phi_0)}
\delta(0),
\label{densi}
\end{equation}
because of the DE-DM coupling; here the subscript $o$ indicates values
at the present time $\tau_{o}$ (it is $a_{o}=1$). We can then assign to
each DM particle a varying mass $M_c (\phi)=M_{oc}e^{-C (\phi -\phi_0)}$

Then, owing to eq.~(\ref{densi}), and assuming that the 
density of the particle widely exceeds the background density, it is
\begin{equation}
\Omega_c \delta_c = {\rho_{M_{c}} - \bar \rho _{c} \over \rho _{cr}}
={ 8 \pi G \over 3 {\mathcal{H}}^2 a} M_{c}(\phi) \delta (0) ~,~~
\Omega_b \delta_b = 
{\rho_{M_{b}}- \bar \rho_b \over \rho_{cr}}
={8\pi G \over 3{\mathcal{H}}^2 a}  M_b \delta (0),
\end{equation}
($\rho_{cr}$ is the critical density and $\delta $ is the Dirac
distribution). Reminding that $\nabla \cdot {\bf v}_{c,b}=
\theta^{c,b} {\mathcal{H}}$ and using the ordinary
(not conformal) time, eq.~(\ref{delta1}) yields
\begin{equation}
\nabla \cdot \dot {\bf v}_c=
- H (1-2\beta X)\, \nabla \cdot {\bf v}_c - 4\pi G a^{-2}
\left( \gamma M_{c}(\phi)  + M_{b} \right) \delta (0)
\label{velocity2}
\end{equation}
(dots yield differentiation in respect to ordinary time and
$H=\dot{a}/a$). Taking into account that 
the acceleration is radial, as the attracting particles lie at the origin,
it will be
\[
\int d^{3}r \, \nabla \cdot {\dot{\textbf {v}}}
=4\pi \int dr~ {d(r^2 \dot{v})/dr}=4\pi r^{2} \dot{v}.
\]
Accordingly, the radial acceleration of a DM particle read
\begin{equation}
\dot{v}_{c}=-(1-2\beta X)H{\textbf {v}}_{c}\cdot 
{\textbf {n}}-\frac{G^{*}M_{c}(\phi)}{R^{2}}-\frac{GM_{b}}{R^{2}},
\label{finalc}
\end{equation}
($\textbf {n}$ is a unit vector in the radial direction; $R=ar$). 

Repeating the calculation for a baryon we get immediately the result
\begin{equation}
\dot{v}_{b}=-H{\textbf {v}}_{b}\cdot 
{\textbf {n}}-{\frac{GM_{c}(\phi)}
{R^{2}}}-{\frac{GM_{b}}{R^{2}}}
\label{finalb}
\end{equation}
In the presence of a full spherical symmetry, $\bf {v}_{b,c} \cdot 
{n}$ $= v_{b,c}$ and using $b=r_b$ and $c=r_c$,  it is easy to derive
eqs.~(\ref{shells}).

For the sake of completeness, here below, we write 
also the equations for the physical DM and baryons radii which easily follow 
from eqs.~(\ref{finalc}) and (\ref{finalb}). 
While for baryons we have the usual Friedman-like equations
\begin{equation}
\ddot R^b_n = - {4\pi \over 3}G ~[\rho_c + \rho_b + \rho_{DE} (1+3w)]~R^b_n ~,
\end{equation}
for DM we have
\begin{equation}
  \ddot R^c_n = C \dot \phi \dot R^c_n - C \dot \phi H R^c_n
- {4\pi \over 3}G ~[\bar \rho_c + \gamma (\rho_c - \bar \rho_c)~ 
+ \rho_b + \rho_{DE} (1+3w)]~R^c_n.
\end{equation}

\section{Discussion}
In this work we studied the evolution of a top--hat fluctuation,
assuming it spherical and that its growth proceeds without external
perturbations.  Neither assumption is realistic. In spite of that, it
is known that the results of a spherical unperturbed growth, obtained
considering only one spatial coordinate, yield a significant insight
into the evolution of fluctuations in the 3--dimensional space.

It seems clear, {\it e.g.}, that baryon outflow from original
density fluctuations is not peculiar for spherical overdensities.  The
fraction of baryons laying outside, at the various growth stages, on
the contrary, can only be considered as an estimate.

If external perturbations are completely negligible, the general
outcome is that the final virialized system is however richer of
DM. This is true indipendently of the way how virialization is
defined. If we define it referring to DM and inner baryons only, the
contribution of the outer baryon layers is clearly missing. But, even
if we wait and define virialization when all baryons, initially
belongong to the fluctuations, have accreted, a substantial amount of
DM, previously external to the top--hat, has also accreted by then. In
both cases the final DM/baryon ratio has apparently increased and this
is true also if we stop at any intermediate stage.

If we look closer at the quantitative details, it appears that the DM
excess is greatest when we refer to DM and inner baryons only and
becomes gradually smaller when we include further layers.  The actual
DM/baryon ratio of real systems will therefore depend on the stage up
to which we take accretion into account. In turn, this is linked to
the second assumption, that the growth occurs without external
perturbations. Steady accretion can be assumed to end when
perturbations can no longer be neglected. Let us return on this point
below.

It would however be fallacious to deduce that the DM/baryon ratio,
in all stabilized systems, exceeds cosmological proportions.
When a DM richer system is formed, baryons are not destroyed,
but enrich the cosmic medium. Systems forming later will therefore
start their growth in a baryon richer environment. When their
formation proceeds unperturbed until the last baryon layer has
accreted, the residual DM excess can be smaller than the initial
baryon excess and the system can be even richer of baryons than
cosmic proportions prescribe.

The above arguments seem therefore to indicate that a gradual decrease
of the average DM/baryon ratio will occur when going from smaller to
greater systems. Large system form later and suffer less important
perturbations. In cDE theories, therefore, galaxy clusters can be
expected to preserve or exceed the cosmological baryon content.

Inside large halos, yielding galaxy clusters, smaller subhalos
continue to grow. Inside a cluster, the interaction of systems is
frequent and outer layers are likely to be stripped. We can therefore
expect, in cDE theories, that galaxies in clusters are DM richer,
while a large deal of baryon materials are stripped by tidal forces in
close encounters, presumably form baryon rich filaments and finally
enrich the intracluster medium (ICM).

Observations show that a hot intracluster plasma with temperatures $T
\sim 10^7$--$10^8$K emits over a broad band from EUV to $X$--ray, from
the potential well of galaxy clusters. Most observed radiation
features are accounted assuming that the emission process is thermal
breemstrahlung, plus lines associated with the metals in ICM. Many
observations in EUV \cite{Lieu} and
soft $X$--ray \cite{bonamente}
however claim an excess emission.  Large n--body simulations (see,
e.g., \cite{cheng}) have been used to try to locate the origin of this
excess, which apparently arise from filaments at $T \sim 10^5$--$10^6
$K$\, $. Simulations apparently succeed in explaining the EUV and soft
X--ray excess, but fail to locate it in filaments.  In turn, filaments
could be a natural product of outer baryon stripping.

It is clearly premature to suggest that baryon--DM segregation helps
to reach an agreement between data and simulation outputs.  There are
however good grounds to assess that the whole picture of ICM emission
is modified by segregation.

The existence of a large amount of baryons in ICM is also testified by
optical data based, {\it e.g.}, on planetary nebulae observations
(see, {\it e.g.}, \cite{arnaboldi} and references therein). Again,
simulations meet data with some reserves (see, e.g., \cite{murante}
and references therein). Clearly, baryon stripping from forming
sub--halos would substantially modify model predictions. In
particular, stars in ICM could turn out to be not so older than the
average galactic populations.

When the smallest galactic systems are then considered, a clear
predictions of cDE theories is that they can be significantly baryon
poor. Another prediction is that their individual baryon contents will
be variable, depending on the frequency and timing of close encounters
with greater structures. It seems however rather clear that a galaxy
satellite, before being dynamically settled, can easily meet bigger
objects and suffer tidal stripping of outer layers.

In the recent years there has been a wide discussion on observed
galaxy satellites. \cite{klypin1999} and \cite{Moore99} showed that,
in a $\Lambda$CDM cosmology, the number of predicted satellite
galaxies exceeds the observed number in the local group by more than
one order of magnitude.  Predictions for $\Lambda$CDM are confirmed
also for dynamical DE \cite{klypin2003}.  Various mechanisms were then
proposed, by a number of authors, to reduce the abundance of
substructure \cite{satelliti1} or to suppress star formation in small
clumps via astrophysical mechanisms, making them dark
\cite{satelliti2}.

Because substructure can modify the fluxes of lensed images relative
to those predicted by smooth lens models, gravitational lensing can be
used to detect and constrain populations of DM clumps in galactic
halos \cite{satelliti3}.  Different methods are, however, proposed by
several authors \cite{satelliti4}.

We cannot state here that the baryon depletion of very small halos,
envisaged here, is the solution of the above problem.  The question
however deserves to be more deeply examined through suitable
techniques.

\begin{figure}
\begin{center}
\vskip -.1truecm
\includegraphics*[width=11cm]{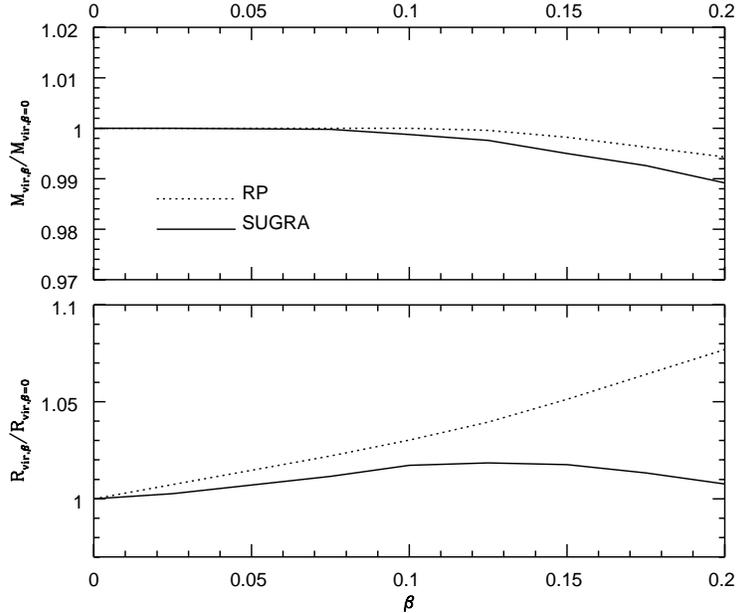}
\end{center}
\vskip -1.2truecm
\caption{Upper panel: Dependence on $\beta$ of the virial mass.
Lower panel: Dependence  on $\beta$ of the virialization radius.}
\vskip .1truecm
\label{masse}
\end{figure}

\section{Conclusions}
In this work we studied the spherical growth of top--hat fluctuations
in cDE theories. Our findings concern: (i) The ambiguity of the
definition of halo virialization, within such context which, in turn,
could cause some difficulty in comparing simulations outputs or data
with Press \& Schechter or similar predictions. (ii) The baryon--DM
segregation during the spherical growth. 

The (i) effect is critical if one wants to use results on the
gravitational growth of a spherical fluctuation to perform predictions
on the mass function for large bound systems, like galaxy clusters.
In fact, the mass enclosed in the virialized region, indipendently
of its definition, does not coincide with the mass enclosed in a
fluctuation evolving according to the linear Jeans' equations.

In upper panel of Figure \ref{masse} we give the ratio between the
virialized mass for various $\beta $ values and the virialized mass for
$\beta=0$, coinciding with the mass in a linearly evolving
fluctuation. The shift arises both from a different virial density
contrast (see Fig.~\ref{Dvir}) and a different virial radius
(Figure~\ref{masse}, lower panel).

The construction of a cluster mass function should take into account these
outputs. In a forthcoming paper, we shall deal with this question in full detail.

Various consequences of the (ii) effect, on different scales, were
also envisaged. On very large scales it leads to predict some baryon
enrichments of large clusters. On intermediate scales it does
interfere with a number of tentative matches between simulations and
observations. On small scales it could offer a way out to the empasse
arising from the scarsity of observed galaxy satellites in the Local
Group.

Quite in general, this analysis shows that cDE theories can open
interesting perspectives in the treatment of a number of cosmological
problems. In turn, observations can be expected to put precise
limits on the strength of the possible DM--DE coupling.

\begin{acknowledgments}
\section{Acknowledgments}
It is a pleasure to thank Silvio Bonometto for his help and useful
discussion which led improvements in the paper. I also thank Luca
Amendola, Loris Colombo and Claudia Quercellini for their comments on
this work.
\end{acknowledgments}

{}


\begin{thebibliography}{}

\bibitem{sugra}
Wetterich C. 1988, Nucl.Phys.B 302, 668

\bibitem{RP} 
 Ratra B. \& Peebles P.J.E., 1988, Phys.Rev.D 37, 3406

\bibitem{PR} 
 Peebles P.J.E \& Ratra B., 2003, Rev. Mod. Phys., 75, 559

\bibitem{Amendola1999}
Amendola L., 1999, Phys. Rev. D60, 043501

\bibitem{lucaclaudia}
Amendola L. \& Quercellini C., 2003, Phys. Rev. D69


\bibitem{maccio}
Maccio' A. V., Quercellini C., Mainini R., Amendola L., Bonometto S. A., 
2004 Phys. Rev. D69, 123516

\bibitem{cmb}
Mainini R., Colombo L. \& Bonometto S.A., 2005, ApJ accepted,  astro-ph/0503036

\bibitem{brax}
Brax, P. \& Martin, J., 1999, Phys.Lett., B468, 40; 
Brax, P. \& Martin, J., 2001, Phys.Rev. D61, 10350;
Brax P., Martin J., Riazuelo A., 2000, Phys.Rev. D62, 103505

\bibitem{PS} 
Press W.H. \& Schechter P.,  1974, ApJ, 187, 425

\bibitem{ST}
Sheth R.K. \& Tormen G., 1999 MNRAS, 308, 119;
Sheth R.K. \& Tormen G., 2002 MNRAS 329, 61;
Jenkins, A., Frenk C.S., White S.D.M., Colberg J.M.,
Cole S., Evrard A.E., Couchman H.M.P. \& Yoshida N., 2001, MNRAS, 321, 372

\bibitem{Lahav_brian}
Lahav, O., Lilje, P.R., Primack, J.R. \& Rees, M., 1991, 
MNRAS, 282, 263E;
Brian, G. \& Norman, M., 1998, ApJ, 495, 80

\bibitem{io}
Mainini R., Maccio' A. V., Bonometto S. A., 2003, New Astron., 8, 173
Mainini R., Maccio' A. V., Bonometto S. A., Klypin A., 2003, ApJ, 599, 24

\bibitem{collapse1} 
Wang L. \& Steinhardt P.J., 1998, ApJ, 508, 483;
Lokas E. L., Bode P., Hoffman Y., 2004, MNRAS, 349, 595
Horellou C., Berge J., 2005, astro-ph/0504465 
Nunes N. J., da Silva A. C., Aghanim N., 2005, astro-ph/0506043

\bibitem{collapse2}
Nunes N. J.\&  Mota D.F, 2005, astro-ph/0409481
Manera M. \& Mota D., 2005, astro-ph/0504519

\bibitem{Amendola03}
Amendola L., Phys. Rev. 2004, D69, 2004, 103524 

\bibitem{maor_wang}
Maor I., Lahav O., 2005, astro-ph/0505308;
Wang P., 2005, astro-ph/0507195; 
Mota D. \& van de Bruck C., 2004, A\&A, 421,71

\bibitem{Lieu} 
Lieu, R., Mittaz,J.P.D., Bowyer, S. et al., 1996 ApJ,
458, L5 and 1996 Science, 274, 1335;
Mittaz, J.P.D., Lieu, R., Lockman, F.J. 1998, ApJ, 498, L17;
Maloney, P.R.,\& Bland-Hawthorn, J. 2001, ApJ, 553, L129

\bibitem{bonamente} 
Bonamente, M., Lieu, R. \& Mittaz, J.P.D. 2001,
ApJ, 547, L7;
Bonamente, M., Nevalainen, J., \& Kaastra,
J.S. 2001, ApJ, 552, L7;
Finoguenov, A., Briel,G. U., Henry, P. J. 2003, A\&A, 410, 777;
Kaastra, J. S., Lieu, R., Tamura, T., Paerels, F. B. S. \& den
Herder,J. W. 2003, A\&A, 397, 445

\bibitem{cheng}
Cheng L.-M., Borgani S., Tozzi P., Tornatore T., Diaferio A., 
Dolag K., He X.-T., Moscardini L., Murante G., Tormen G., 2005,
astro-ph/0409759, to appear in A\&A

\bibitem{arnaboldi}
Gerhard O., Arnaboldi M., Freeman K.C., Kashikawa N., Okamura S.,  Yasuda N.,
 Astrophys.J. 621 (2005) L93

\bibitem{murante} 
Murante G., Arnaboldi M., Gerhard O., Borgani S.,
Cheng L.M., Diaferio A., Dolag K., Tornatore L., Tozzi P.,
Astrophys.J. 607 (2004) L83-L86

\bibitem{klypin1999}
  Klypin A., Gottloeber S., Kravtsov A. \& Khokhlov A., 1999 
  ApJ 516, 530

\bibitem{Moore99} Moore, B., Ghigna, S., 
Governato, F., Lake, G., Quinn, T., Stadel, J., \& Tozzi, P.\ 1999, ApJ Lett., 
524, L19 

\bibitem{klypin2003}
  Klypin A., Macci\`o A., Mainini R. \& Bonometto S.A., 2003
  ApJ  599, 31

\bibitem{satelliti1}
Spergel, D.~N., \& Steinhardt, P.~J. 2000, Phys. Rev. Lett., 84, 3760;
Hannestad, S., \& Scherrer, R.~J. 2000, Phys. Rev. D, 62, 43522;
Hu, W., Barkana, R., \& Gruzinov, A. 2000, Phys. Rev. Lett, 85, 1158

\bibitem{satelliti2}
Bullock, J.~S., Kravtsov, A.~V., \& Weinberg, D.~H.\ 2000, ApJ, 539, 517;
Benson, A.~J., Lacey, 
C.~G., Baugh, C.~M., Cole, S., \& Frenk, C.~S.\ 2002, MNRAS, 333, 156;
Somerville, R.~S.\ 2002, ApJ Lett, 572, L23 

\bibitem{satelliti3}
Mao, S., \& Schneider, P. 1998, MNRAS, 295, 587;
Metcalf, R.~B., \& Madau, P. 2001, ApJ, 563, 9;
Metcalf, R.~B., \& Zhao, H. 2002, ApJ Lett., 567, L5;
Dalal, N., \& Kochanek, C.~S. 2002, ApJ, 572, 25

\bibitem{satelliti4}
Mayer, L., Moore, B., Quinn, T., Governato, F., \& Stadel, J. 2002,
  MNRAS, 336, 119;
Bergstr{\" o}m L., Edsj{\" o} J., Gondolo P., \& Ullio P. 1999,
  Phys. Rev. D59, 43506;
Tasitsiomi A., \& Olinto A.~V. 2002, Phys. Rev. D66, 83006


\end{thebibliography}
\end{document}